# Manipulation of conformational change in proteins by single residue perturbations

C Atilgan, Z N Gerek, S B Ozkan, A R Atilgan

**ABSTRACT**

Using the perturbation-response scanning (PRS) technique, we study a set of 23 proteins that display a variety of conformational motions upon ligand binding (e.g. shear, hinge, allosteric). In most cases, PRS determines residues that may be manipulated to achieve the resulting conformational change. PRS reveals that for some proteins, binding induced conformational change may be achieved through the perturbation of residues scattered throughout the protein, whereas in others, perturbation of specific residues confined to a highly specific region are necessary. Correlations between the experimental and calculated atomic displacements are always better or equivalent to those obtained from a modal analysis of elastic network models. Furthermore, best correlations obtained by the latter approach do not always appear in the most collective modes. We show that success of the modal analysis depends on the lack of redundant paths that exist in the protein. PRS thus demonstrates that several relevant modes may simultaneously be induced by perturbing a single select residue on the protein. We also illustrate the biological relevance of applying PRS on the GroEL and ADK structures in detail, where we show that the residues whose perturbation lead to the precise conformational changes usually correspond to those experimentally determined to be functionally important.

**INTRODUCTION**

A comparison of the experimentally determined *apo* and *holo* forms of a protein provides a wealth of information on the basic motions involved upon binding of the ligand, as well as the residues participating in functionality. Such knowledge base is extremely important for deciphering key residues that may be targeted for drug design purposes, or enhancing enzymatic activity. Computational techniques have been developed to distill information from the structures, going beyond the original studies that relied on their visual inspection. Normal mode analysis, in particular, turned out to be a useful technique in the analyses of functionally related conformational dynamics (1-3). As such, it has been possible to classify protein motions, e.g. hinge bending, shear (4). It has further been shown that the predominant contributions to these motions may be described by a single, most collective mode for some proteins, whereas it may be obtained from a superposition of several modes for others (5). With the advent of coarse graining of biomolecular structures through residue-based network models (4, 6-13), it has been possible to study a large number of protein structures. These anisotropic network models (ANMs) take into account the three dimensional geometry of interacting pairs of residues to study the modal behavior of proteins. Using such information, it is possible to morph between the apo and holo structures to gain insight towards the intermediates that lead to the final structure (14-17). These analyses also uncover the different modes stimulated by different ligands binding to the same apo form. Moreover, displacement vectors obtained from the modes might be used to generate alternative starting structures for molecular dynamics (MD) simulations, or to make educated guesses for missing residues in the experimentally determined structures (18). In addition, docking algorithms are improved by incorporating the flexibility of receptor in the calculations through the normal mode predictions (19-21).

Though there is considerable success in these classifications, there is also a debate, and lack of insight, on the origin of how these modes are utilized by different types of proteins, or different ligands acting on the same protein. This problem has been addressed by a methodology that assesses the number of modes necessary to map a given conformational change (18). Therein, the degree of accuracy obtained from the inclusion of a given number of modes was shown to be protein dependent. In another study where 170 pairs of structures were systematically analyzed, it was shown that the success of coarse-grained elastic network models may be improved by recognizing the rigidity of some residue clusters (22). These authors further demonstrated that the collectivity of the motion is detrimental in the ability to represent the motion by a few slow modes.

Albeit its success in identifying the types of motions involved in ligand binding, ANM is an equilibrium methodology, providing information on the equilibrium fluctuations and the various contributions to those fluctuations from different modes of motion. Similar to experimental procedures that lead to useful information, it is of interest to devise a methodology where a perturbation is inserted into the system so as to displace the structure slightly out of equilibrium. The response is recorded to detect the underlying features contributing to the observations, yielding valuable information, beyond the correlations between the fluctuations inherent in the system at equilibrium.

In the COREX algorithm, such perturbations are introduced in terms of folded-unfolded subunits in an ensemble of the protein, whereby the high resolution experimental structure is used as a template (23). The effect of environmental variables such as the temperature or mutations on the protein are reproduced by this methodology; it is also used in predicting cooperativity and long-range communication in proteins (24). A perturbation–response technique applied to the all-atom model of the protein was used to elucidate the shifts in the energy landscape accompanying binding (25). Therein, the perturbations are introduced as local displacements of selected atoms followed by energy minimization; then the response is measured as the relative ability of a residue to induce displacements on other residues, versus its propensity to resist change. This molecular mechanics based methodology relies on scanning all the residues to produce comparative results (26). In another near equilibrium methodology, this time as applied to the coarse-grained network structure of a protein, one can insert forces in the contact between a chosen pair of residues and uncover the control mechanism that adaptively annihilates the induced change (27). Similar approaches have been adopted in other work, whereby the perturbations on residues are introduced by modifying the effective force constants (28), distances (29, 30) between contacting pairs or both (31). Conversely, one may insert the forces on the nodes instead of the links between pairs of nodes; depending on the location of the perturbation, the resulting displacements between the apo and holo forms may be highly correlated with those determined experimentally (32). Recently, this approach was extended to scan the whole protein residue-by-residue, a methodology termed perturbation-response scanning (PRS) (33). By recording the response to each inserted force on the ferric binding protein (FBP), it has been possible to map those residues that are structurally amenable to induce the necessary conformational change upon binding. Moreover, independent of the directionality of the inserted force in a remotely located residue, the directionality of the response is found to be organized around the binding region of FBP. The strength of all the above-mentioned techniques is that, they recognize the contributions to the motions from all degrees of freedom of the system, rather than focusing on a subset represented by a few slow normal modes. These methods are able to bring out an underlying feature of the system by observing the response to a disturbance of the system slightly out of equilibrium, within the linear-response regime.

We study 23 pairs of structures using PRS and ANM. We show that PRS maps residues that may alone manipulate the structure between the apo – holo forms during ligand binding. With ANM, we determine the mode whose base vector best reproduces the motion between the two forms for each protein pair. We show that the manipulation of a single residue will reproduce the structural change better than that of the best mode in ANM in all the cases inspected. The variability of the number and location of the residues that reproduce the displacement profile lends clues on how that particular protein functions, as shown by case studies.

**METHODS**

***Theory.*** Here we present a review of how a native protein structure may be manipulated by external forces (33). We assemble the folded protein as a network of $N$ nodes that are on the $C_\alpha$ atoms. Any given pair of nodes interacts in accord with a conventional harmonic potential, if the nodes are within a cut-off distance $R_c$ of each other.

In the notation used, **r** and **f** are the bond and internal force vectors along the edge connecting any two nodes, respectively. On the other hand, **R** and **F** are vectors on the nodes and are called the position and external force vectors, respectively. There are $m_i$ interactions for each residue $i$ (e.g. residue 11 may have four interactions, $m_{11} = 4$), and a total of $M$ interactions for $N$ residues (*i.e.*, $M = \Sigma\, m_i/2$). In the absence of external force acting on the system, the equilibrium condition for each residue, $i$, requires that the summation of the internal, residue-residue interaction forces must be zero for each residue:

$$\mathbf{b}\Delta\mathbf{f}_i = 0 \qquad (1)$$

where the $3 \times m_i$ coefficient matrix **b** consists of the direction cosines of each force representing the residue-residue interaction. The row indices of **b** are $x$, $y$, or $z$. Here $\Delta\mathbf{f}_i$ is an $m \times 1$ column matrix of forces aligned in the direction of the bond between the two interacting residues. For instance, if a residue $i$ has nine contacts, $\Delta \mathbf{f}_i$ is a $9 \times 1$ column matrix. Following the numerical example, equation 1 sums up the projection of these nine forces on the $x$, $y$, and $z$-axes.

One can write the equilibrium condition (eq 1) for each residue. This leads to a total of $N$ sets of equations, each of which involves the summation of forces in three respective directions. Generalizing eq 1 to the whole system of $N$ residues and $M$ interactions, one can write the following algebraic system of a total of $3N$ equations consisting of $M$ unknown residue-residue interaction forces

$$\mathbf{B}\Delta \mathbf{f}_i = 0 \tag{2}$$

with the $3N \times M$ direction cosine matrix $\mathbf{B}$ and the $M \times 1$ column matrix of residue-residue interaction forces, $\Delta \mathbf{f}$. It is straightforward to generate the matrix $\mathbf{B}$ from the topology of the native structure (*i.e.*, the protein data bank (PDB) file (34)) with a specified $r_c$. As an example, Formate Dehydrogenase (FD) has 374 residues and a total number of 1816 interactions when $R_c = 8.0$ Å is selected.

In the presence of an external force, $\Delta \mathbf{F}$, the equilibrium condition for each residue imposes that the summation of the residue-residue interaction forces must be equal to the external, applied force on the same residue. Then, eq 2 may be expressed as

$$\mathbf{B}_{3N \times M} \Delta \mathbf{f}_{M \times 1} = \Delta \mathbf{F}_{3N \times 1} \tag{3}$$

Under the action of external forces, each residue experiences a displacement, $\Delta \mathbf{R}$, which is the positional displacement vector. Moreover, the bond distance between any two residues changes by $\Delta \mathbf{r}$ in accord with the positional displacements of two residues which participate in the contact. Therefore, there must be compatibility between the $3N$ positional displacements and the changes that take place in the intra-residual distances, a total of $M$ distortions. This compatibility is very similar to the form given in eq 3 (27):

$$\mathbf{B}^T_{M \times 3N} \Delta \mathbf{R}_{3N \times 1} = \Delta \mathbf{r}_{M \times 1} \tag{4}$$

Within the scope of an elastic network of residues that are connected to their neighbors by linear-elastic springs, the residual interaction forces, $\Delta \mathbf{f}$, are related to the changes in the contact distances, $\Delta \mathbf{r}$, through Hooke's law by

$$\mathbf{K}_{M \times M} \Delta \mathbf{r}_{M \times 1} = \Delta \mathbf{f}_{M \times 1} \tag{5}$$

where the coefficient matrix $\mathbf{K}$ is diagonal. We take the entries of $\mathbf{K}$ to be equivalent. We have previously validated this assumption by comparing residue cross-correlations obtained from the simplified Hookean potential in eq 5 with those from MD (25, 26, 33).

Thus, rearranging eq 3-5, one gets the forces necessary to induce a given point-by-point displacement of residues:

$$\left(\mathbf{B}\mathbf{K}\mathbf{B}^T\right)\Delta \mathbf{R} = \Delta \mathbf{F} \tag{6}$$

On the other hand, one may choose to perturb a single or a set of residues, and follow the response of the residue network through,

$$\left(\mathbf{B}\mathbf{K}\mathbf{B}^T\right)^{-1} \Delta \mathbf{F} = \Delta \mathbf{R} \tag{7}$$

where the $\Delta \mathbf{F}$ vector will contain the components of the externally applied force vectors on the selected residues.

***Perturbation-response scanning.*** We analyze a set of 23 proteins in apo and holo forms (Table 1). Unless otherwise specified, for each pair of experimental structure, the holo is superimposed on the apo form, followed by the computation of the residue displacement vectors, $\Delta\mathbf{D}$. In PRS, we perturb the apo form of each protein by applying a force to the $C_\alpha$ atom of a residue. We scan the protein, consecutively perturbing each residue, $i$ by applying the force $\Delta\mathbf{F}$. Thus, the elements of the $\Delta\mathbf{F}$ vector are nonzero only for the three terms ($\Delta F_{3i-2}$, $\Delta F_{3i-1}$, $\Delta F_{3i}$). We then record the expected changes, $\Delta\mathbf{R}$, as a result of the linear response of the protein, computed through eq 7. We report the averages over ten independent runs where the forces are applied randomly in each direction, with the magnitudes chosen uniformly in the interval [-0.1, 0.1]. One realization of PRS on the largest system studied, topoisomerase II (TII) with $M = 8877$ interactions, takes 30 min time on Intel Xeon 2.70 GHz CPU. The smallest systems take ca. 10 seconds to analyze.

To assess the quality of the predicted displacements of all residues resulting from a force applied on selected residue $i$, we use two metrics. The first is the Pearson correlation between the elements of the $\Delta\mathbf{R}$ and $\Delta\mathbf{D}$ vectors. We define $(\Delta R_k)^i$ as the theoretical predictions of the size of the displacement of each residue, $k$, as a response of the system to applied forces on residue $i$ (eq 7). $\Delta D_k$ are the size of the displacements between the apo and the holo forms obtained from the PDB structures. Thus, $\Delta R_i^k$ are compared with $\Delta D_k$ and the goodness of the prediction is quantified as the Pearson correlation coefficient for each perturbed residue $i$:

$$C_i = \frac{\sum_{j=1}^{N}\left[(\Delta R_k)^i - (\overline{\Delta R})^i\right](\Delta D_k - \overline{\Delta D})}{(N-1)\,\sigma_R \sigma_D} \qquad (8)$$

where the overbar indicates the average. $\sigma_R$ and $\sigma_D$ are the respective standard deviations of calculated and experimental displacements.

While $C_i$ is a measure of the goodness of fit of the size of the displacements, that of the directionality of the response may be quantified by the overlap of the predicted and experimental displacements

$$O_i = \frac{(\Delta\mathbf{R})^i \cdot \Delta\mathbf{D}}{\left|(\Delta\mathbf{R}\cdot\Delta\mathbf{R}^T)^i (\Delta\mathbf{D}\cdot\Delta\mathbf{D}^T)^i\right|^{1/2}} \qquad (9)$$

For both metrics, a value close to 1 implies good agreement with experiment, and lack of correlation between experiment and theoretical findings yields a value of zero.

***Modal analysis.*** Equivalence of the equilibrium fluctuations obtained from the normal modes and the force balance (eq 6) was previously shown, wherein the modal decomposition was also performed (6). Thus, ($\mathbf{BKB^T}$) is the $3N \times 3N$ matrix equivalent to the Hessian, $\mathbf{H}$, of the system studied. The pseudo-inverse of the $\mathbf{H}$ matrix is obtained as

$$\mathbf{H}^{-1} = \mathbf{U}\,\mathbf{\Lambda}^{-1}\,\mathbf{U}^T \qquad (10)$$

$\mathbf{\Lambda}$ is a diagonal matrix whose elements $\lambda_j$ are the eigenvalues of $\mathbf{H}$, and $\mathbf{U}$ is the orthonormal matrix whose columns $\mathbf{u}_j$ are the eigenvectors of $\mathbf{H}$. $\mathbf{H}$ has at least six zero eigenvalues corresponding to the purely translational and rotational motions. In modal analysis, for a given mode $j$, $\mathbf{u}_j$ are treated as displacement vectors. Pearson correlation and overlap (eq 8 and 9) between the $3N$ elements of the $\mathbf{u}_j$ and $\Delta\mathbf{D}$ vectors are used to select the mode, $j$, which best describes the binding motion.

To assess the quality of the modes obtained by $\mathbf{H}^{-1}$, we use the same two metrics as for PRS;

(i) Pearson correlation of mode $i$ with the observed conformational changes using eq 8, and (ii) similarity between the direction of $i^{th}$ mode and observed conformational change of the protein based on overlap equation (eq 9). In both metrics, the displacement vector upon perturbation, $\Delta \mathbf{R}$ is replaced with the normal vector, $\mathbf{u}_j$.

***Optimization of the cut-off distance.*** The eigenvalue distribution of the Hessian of proteins is such that the low frequency region is more crowded than expected of polymers or other condensed matter (35). Thus, the choice of the cutoff distance, $R_c$, for the construction of the Hessian is critical for extracting protein-like properties from the systems studied. For all the proteins studied in this work, we coarse-grain the PDB structure so that each residue is represented by the coordinates of its $C_\alpha$ atom. We then repeat the PRS analysis for a variety of cut-off distances in the range of 7.0 – 14.0 Å in increments of 0.1 Å; the lower limit defines the first coordination shell of residues in proteins. For each network structure, we ensure that the system has six zero eigenvalues corresponding to the translational and rotational degrees of freedom of the protein. For a given protein we select the cut-off distance, $R_c^{opt}$ that yields the closest agreement of the displacement vectors from experiments for at least one residue. We verify that the correlation and overlap values reported in the **Results** section are not affected for a range of values between $R_c^{opt} \pm 0.2$ Å for all the 23 proteins studied. We also verify that the order of residue indices that provide the best correlations/overlaps do not change within this range of $R_c^{opt}$. Note that $R_c^{opt}$ is independent of the size of the protein (Table 1). Optimization is done separately for PRS and ANM.

**RESULTS**

***Analysis of proteins with different conformational change classifications.*** We study the conformational change upon ligand binding of a set of 23 proteins (Table 1) that show various types of motions such as shear, hinge, allosteric, partial refolding and more complex motions within subdomains, between subdomains and between subunits of the proteins as classified in the Yale Morph server (14). For each protein we perform two analyses:

(i) With PRS, we scan the protein by inserting random forces on all residues sequentially. For each residue, we then record the overlap of the response vector with the experimental conformational change vector ($O_i$, eq 9) and correlation coefficients between the size of the displacements of the calculated and experimental conformational changes ($C_i$, eq 8). In Table 1, we report the highest overlap and correlation coefficient values from the average over the $m = 10$ runs as well as the single best case obtained.

(ii) With ANM, we seek the mode that best represents the conformational change. We therefore calculate the overlap and correlation coefficients of each eigenvector of the Hessian matrix ($\mathbf{u}_j$) with the experimentally determined conformational change between the apo and holo structures ($\Delta \mathbf{D}$). The highest overlap and correlation values are reported in the last two columns of Table 1 along with the mode number that reproduces this result shown in parentheses. Note that modes are sequentially numbered from the slowest to the fastest, excluding the six corresponding to the translation and rotation of the protein; therefore, the most collective non-trivial mode is numbered as 1.

We first observe from the PRS results that there are a few proteins for which the perturbation of a single residue reproduces both the direction and the magnitude of the conformational change pattern, as measured by overlap and Pearson correlations (setting a threshold of 0.7 or better, shaded gray in Table 1). These apo and holo structures are linearly connected, although the size of the conformational change may be relatively large (RMSD between the structures are in the range 1.2 – 12.4 Å). These include GroEL that have complex motions of the subunits involved in the conformational change, as well as all four of the proteins displaying hinge motions between their domains, i.e. adenylate kinase (ADK), FD, ferric binding protein (FBP), maltose binding protein (MBP). Of these, we will study ADK and GroEL in detail in subsection C. FBP was studied in detail previously using the PRS technique and MD (33).

Further, for 20 of the 23 proteins, the relative magnitudes of the residue displacements may be well-reproduced by singly placed forces on select residues. Except for molybdate binding protein, immunoglobulin, and HIV-1 reverse transcriptase, at least one residue may be found to carry out such a manipulation. Large overlaps are obtained less, indicating that although the size of the response may be captured, the exact direction of the response is more difficult to achieve by such single residue manipulations.

Interestingly, although many of these proteins display large overlap and/or correlation values, the numbers of residues that lead to them differ between proteins. For some proteins there is not much specificity on the residue to be perturbed to reproduce the conformational change. For others, by perturbing a very specific location, the whole conformational change is obtained. The former is exemplified by FBP: As long as one avoids perturbations on deeply buried residues or those that reside in key locations on secondary structural units, a force exerted on single residues lead to the conformational change (33). In such proteins, the holo form is thought to reside close to the apo form on the free energy surface, as a weakly populated conformation.

Our observations from the ANM analysis seeking a single mode that best represents the conformational change have different qualities. Similar to the PRS analysis, three of the four proteins that have hinge motions between the domains (ADK, FBP and MBP) along with GroEL display high overlap and correlations with the experimentally measured conformational change upon binding. In all these cases, the conformational change is best represented by one of the two slowest modes. In none of the other proteins is it possible to achieve high overlaps by projecting the motion on a single mode, while thymidylate syntase, tetracycline repressor (TR), ubiquitin conjugating enzyme (UCE), cytochrome P450BM-3, aldose reductase, serpin, hemoglobin and TII yield high correlations with the displacements. However, the mode that gives the best representation of the motion need not be one of the most collective in these proteins (e.g. the shear motions between the subdomains of TR motions are reproduced with mode 16 and the complex motions of TII with mode 18).

In fig 1, we display the experimental displacement profiles for four sample proteins, ADK, aspartate receptor (AR), UCE and FD, and we compare them with the best PRS and ANM predictions. We find that PRS outperforms ANM predictions in each case, capturing most of the details recorded in the experimental conformational change. Thus, by perturbing specific residues, it is possible to act on multiple modes so as to induce conformational changes that are functionally relevant. These findings are in accord with studies where the success of single ANM modes to predict conformational change is shown to be highly protein dependent (18). It has been proposed that such dominance of a single mode is related to the collectivity of the transition (22). In the next subsection we use a metric, called redundancy index that specifies the degree of such collectivity, hence the dominance of a single mode.

***Redundancy index as a first order measure of collectivity in the protein.*** Collectivity of motions in a protein depends on the propensity of its residues to find alternative routes to communicate with function-related destinations such as the active site. Our working hypothesis is that the larger the number of pathways between two regions in the protein, the more complicated the motions within the structure are expected to be. Thus, a superposition of a larger number of collective modes shall lead to those motions.

In what follows, we define a "non-bonded contact" between a pair of residues $p$ and $q$ as follows: A residue $q$ within the distance $R_c$ of $p$ is not its first or second neighbor along the contour of the chain; i.e., $|p - q| \geq 3$. We quantify the redundancy in the protein by the number of possible ways to reach a non-bonded contact of a residue, had its direct contact been momentarily screened.

We calculate the number of alternative routes from $p$ to its non-bonded contact $q$ through other edges of $p$ in the residue network. Letting $t_{m,pq}$ be the number of ways of reaching a direct contact $q$ of residue $p$ in $m$ steps, one first forms the *perturbed* adjacency matrix $\mathbf{M}^{pq}$ from the coordinates of the $C_\alpha$ atoms, whose elements are given by

$$M_{ij}^{pq} = \begin{cases} H(R_c - R_{ij}) & i \neq j \\ 0 & i = j \\ 0 & i = p, j = q \end{cases} \quad (11)$$

where $R_{ij}$ is the distance between residues $i$ and $j$, $H(x)$ is the Heaviside step function whose value is zero for $x \leq 0$ and 1 otherwise. Note that eq 11 explicitly states why the matrix $\mathbf{M}$ is referred to as *perturbed*. A selected direct non-bonded contact between the residue $p$ and neighbor $q$ is set to zero. The powers property of the adjacency matrix dictates that its $m^{th}$ power gives the number of $m$-step paths to reach $q$ from $p$. Thus, $t_{m,pq} = (\mathbf{M}^{pq})^m$, and in particular, the number of two-step paths is given by $t_{2,pq} = (\mathbf{M}^{pq})^2$.

The calculation is repeated for all non-bonded contact pairs $(p, q)$. For each residue, $p$, the two-step paths to the non-bonded contacting neighbor $q$ is normalized by the total number of such contacts of the $p^{th}$ residue, $k_p$:

$$t_{2,p} = \frac{\sum_{q=1}^{N} t_{2,pq}}{k_p} \quad (12)$$

The number of redundancies generated by a given residue also depends on its location in the network, the more central residues having higher reachability. This property is quantified by the average shortest path length $L_i$ of the residue which is highly correlated with the experimentally measured residue fluctuations (36). For each residue $i$, it is an average value over the minimum number of connections that must be transversed to connect it with its $j^{th}$ neighbor, $L_i = \frac{1}{N}\sum_{j=1}^{N} L_{ij}$. We employ the Dijkstra algorithm to compute the shortest paths.

Thus, we define the redundancy index, $r$, of the protein as the ratio of the number of alternative two-step paths a given residue $i$ generates to its non-bonded contacts (a local property) and its overall reachability (a global property):

$$r_i = \frac{t_{2,i}}{L_i} \quad (13)$$

In general, one would expect all paths to long-range contacts to be effective in the redundancy index; *i.e.* all screened $n$-step paths, each with $t_{n,i}$. However, the two-step paths are expected to have the largest contributions in the fluctuating environment of the protein when direct paths are momentarily screened.

In fig 2a, we display the relationship between $<r_i>$, averaged over all the residues, and its variance for the set of 23 proteins shown in Table 1. Proteins with a low $<r_i>$ value also display a narrow distribution so that most residues have similar redundancy. Such structures are expected to support collective motions, symbolized by a single or a few slow modes. As examples, in fig 2b, we show the distribution of $r_i$ for ADK and hemoglobin. The average for the former is 0.83±0.21 and it is very narrowly distributed. As a consequence, its conformational change is well-described by the most collective mode ($C_i$ = 0.91 for the slowest mode in Table 1). The latter has a wider distribution with an average value of 1.99±0.59. Its conformational change may still be projected onto a single mode (albeit the third most collective mode) with a somewhat smaller correlation coefficient of 0.79. In fact, by coloring the points in fig 2a by the mode number that best describes the conformational change, we find that the more collective mode-description of the size of the displacement upon binding, $\Delta D$ corresponds to low values of $<r_i>$ with narrow distributions. Conversely, $\Delta D$ best-described by less collective modes are accompanied by very poor overlaps ($O_i$ < 0.4).

We also focus on the contributions to $<r_i>$ from the local and global structure of the protein; i.e., we compare the number of two-step paths of a residue to its long-range contacts, $t_{2,i}$, with its reachability, $L_i$. These are displayed in fig 2c for ADK and hemoglobin. The former has a non-globular structure with $L_i$ in the range 4–10. Although the values of $L_i$ cover a large range, they have a positive correlation with $t_{2,i}$ leading to the narrow distribution of $r_i$ in fig 2b. The behavior observed in the smaller and more globular structure of hemoglobin has the opposite trends in $L_i$ and $t_{2,i}$, leading to a wider distribution of their ratio. Thus, in a given protein structure, there are both centrally and distantly located residues, but the location of a residue need not dictate the number of local alternative connections to the immediate neighbors. It is the balance between these local and global measures that finally determine the collectivity of the motion.

We note that $<r_i>$ is not a perfect measure of the tendency to move collectively, and in many cases the higher order contributions neglected in eq 13 may explain the discrepancies. Nevertheless, it provides a means to inspect a given protein structure to answer if the conformational change can be described by a few normal modes. PRS, on the other hand, poses the alternative question, "Can all the modes that lead to a given conformational change be invoked by perturbing a single residue?" The degree of success presented in Table 1 and fig 1 suggests that the answer is affirmative for many of the proteins studied. Our studies further indicate that those residues that lead to high correlation/overlap values hold key locations in the protein structure. We next inspect two of these proteins in detail to demonstrate how PRS may be utilized to determine structurally/functionally important residues.

***Case Studies Illustrating the Biological Relevance of the PRS method.*** In this subsection, we apply the PRS analysis on two proteins, GroEL and ADK. We choose them as case studies because there are many experimental and computational findings on their conformational dynamics. We investigate if the residues whose perturbation gives highly correlated response vectors with the conformational change upon binding have specific functions determined experimentally. Note that here we perform the PRS analysis on two separate conformations to determine the shift in the roles of residues. Conversely, in Table 1, only the results from open conformations are reported.

**GroEL:** GroEL, a molecular chaperon, helps the folding of substrate proteins that may otherwise aggregate. It uses ATP as the energy source; so binding of ATP to GroEL initiates the allosteric transitions that facilitate the refolding of misfolded substrate proteins. The GroEL macromolecule consists of two heptameric rings stacked back to back. Each subunit of the GroEL has three major domains: the *apical* (residues 191– 376), *equatorial* (residues 1–133, 409–548), and *intermediate* (residues 134–190, 377–408). Each of the two GroEL rings undergoes the same, but out of phase, complex allosteric cycle consisting of a series of conformational changes between T, R, R′, and R″ states. We analyze the first T→R and the last R″→ T transitions by applying PRS on the available crystal structures of the initial and final conformations in these transformations. Following the work of Tehver et al. (37), we employ the PDB structures 1AON (chain H for T state) and 2C7E (chain A for R state) for T→R analysis. Similarly, for R″→T analysis, 1AON (chain A for R″ state) and 1GR5 (chain A for T state) are used.

We investigate the residues playing a critical role for T→R by confining our attention to those which lead to a response vector that is highly correlated with T→R conformational displacement vector. We repeat the same analysis for R″→T transition. Fig 3 shows the ribbon diagram of GroEL (PDB code: 1AON) which is colored by the correlation coefficients between the theoretical and experimental conformational change of T→R transition (A) and R″→T transition (B). The residues with the highest correlation ($C_i > 0.69$) are shown in ball representation. A related analysis based-on an elastic network model (38, 39) that weighs the spring constants by a statistical potential (40) was performed for GroEL allosteric transitions. Interestingly, the hot residues identified with this method and PRS coincide considerably in the allosteric transformation vectors of both T→R and R″→T (the number of residues with $C_i > 0.7$ are 110 and 56, respectively, listed in Table 2).

Furthermore, there are other experimental works which indicate the critical role of most of the identified for T→R and R″→T transitions. For example, based on PRS, we find residues G192 and G375 (shown as part of the set of residues in ball representation in Fig 3) as critical for T→R transition. These are determined as perfectly conserved residues at the transition regions between domains I and A (41). Moreover, another hot residue, G414 is also conserved, undergoing severe backbone torsion between *cis* and *trans* conformations corresponding to the ATP/ADP binding positions (42). In the study of binding potato leafroll virus, R13, K15, L17, and R18 residues of the N-terminal region of the equatorial domain of GroEL are found as critical for binding (43). PRS also indicates that perturbation of these residues is critical to obtain T→R conformational change. Finally, PRS reveals the importance of Cys138 of the intermediate domain for the allosteric changes as corroborated by recent experimental work that shows it affects the structural and functional integrity of the complex (44).

For R″→T transition, hot residues we determine by PRS correspond to the regions 283-289 and 350-369. These contain many charged residues (D283, R284, R285, D359, D361, K364, E367, and R368) that are exposed to the central cavity in the cis-ring. Moreover, the conservation of these residues may be more related to their functional role in interacting with the substrate (42). Similar regions (287-289 and 358-379) were also found in (38). Finally, PRS indicates D359 as playing a critical role in the R″→T transition. A recent experimental study has shown that the mutation of this residue would compromise the ATPase activity (41).

**Adenylate Kinase:** ADK is a phosphotransferase enzyme that catalyzes the conversion of adenine nucleotides, and plays an important role in cellular energy homeostasis. ADK has three major domains: ATP binding (residues 118-167), NMP binding (residues 30-67) and the rest of the structure; we term these LID, NMP, and CORE, respectively. Fig 4 presents the ribbon diagrams of open and closed states of ADK. The perturbed residues that give the highest correlations ($C_i > 0.71$) are shown as balls.

In the case of the apo form, there are many residues whose perturbation leads to displacement vectors, ΔR, highly correlated with the experimental conformational change, ΔD. As shown on Fig 4A, these are clustered in the LID and NMP domains, meaning that perturbing most of the residues in these regions leads to a binding induced conformational change. This agrees with the study based on NMR and MD (45) that there are several meta-stable configurations bridging the open to closed states of ADK. In particular, residues R44 and R149 whose importance in binding have been shown experimentally (46) give high correlation upon perturbing them in the apo conformation ($C_{44} = 0.89$ and $C_{149} = 0.84$, averaged over 10 runs). In the case of the holo form (fig 3B), the number of residues that give high correlations are reduced to a few in the NMP domain (residues 53-57 with $C_i \approx 0.77$) and others in the LID domain (residues 124-133, 144-146, 151-154 with $C_i \approx 0.69$). Thus, the number of residues whose perturbation invokes a response that will make the protein shift from the bound to the unbound conformation is much less in case of holo structure (those with $C_i > 0.7$ is reduced from 120 to only 8; Table 2).

These observations are consistent with the emerging view about ADK binding/catalysis process. As shown experimentally and computationally (47, 48), ADK can exist in a variety of different conformations along the reaction pathway from the apo to holo forms. Further, by mutating several surface exposed LID residues, it was recently shown that the lowest-energy conformation of apo ADK remains unaltered, while the populations of all states are redistributed (49). The control of the binding reaction is achieved by the unbound enzyme conformation, replacing the conventional idea of ligand induced conformational change. Based on this view, ADK populates several conformations that display varying degrees of closeness to the unbound and bound conformation whether or not the ligand is present. The ligand only binds when there is an appropriate partner already in the cleft, followed by the closing of the lid; the reaction then proceeds. Our PRS analysis supports this picture: The unbound conformation of ADK can easily sample the bound conformations, since perturbations on most of the residues in LID and NMP can lead to these conformational changes. On the other hand, once the substrate is bound, the conformational variability of the holo ADK is reduced, with the highest correlation coefficient values being reduced from 0.91 to 0.77. Thus, the substrate stabilizes the holo form, and only by perturbing selected residues in NMP and LID can the apo conformation be regained.

We further investigate why it is easier to get a response vector that is well correlated with the unbound → bound conformational change when we perturb the residues of apo structure of ADK (unbound state). From a topological perspective, this information is contained in $\mathbf{H}^{-1}$ which operates on the exerted forces to yield the response vectors (eq 7). We explore the differences between the magnitude normalized columns of $\mathbf{H}^{-1}$ of the open and closed states (Fig 4C,D). Based on PRS, each normalized column, $i$, in $\mathbf{H}^{-1}$ shows ΔR displacement vector profile of the protein upon applying a unit force in $x$-, $y$-, $z$-direction to residue $i$. Thus, they can be considered as color coded response vector maps. The response vector map of the apo state, unlike that of the holo, displays homogeneity in the columns; (i.e. the color coded map for each column is similar leading to comparable responses). In particular, the predominantly negative values in residues 130–150 that correspond to LID region is noticeably uniform. On the other hand, the heterogeneity in the response vector map of holo state indicates that the responses will depend on the perturbed residue.

**CONCLUSIONS**

By applying PRS on a set of 23 proteins that display a variety of conformational states, we illustrate that for many proteins, one may reproduce the function related conformational changes by forces exerted on selected residues. We show that, although these motions may sometimes be directly related to a single dominant collective mode, for many proteins a superposition of several modes is necessary to achieve the change. Thus, forces exerted on a single residue can induce several modes to operate simultaneously. We further define a metric, called redundancy index, which determines when a single mode can successfully reproduce binding induced motions. It successfully relates the collectivity of motions in a protein to a combination of the local and global descriptors that are relevant to the number of alternative paths for information transfer.

The case studies of GroEL and ADK show that residues yielding displacements highly correlated with the experimental values are also hot spots that were previously implicated experimentally and computationally. Furthermore, in ADK the large number of successful residues in the apo→holo perturbation is related to the conformational multiplicity that is observed in the apo form of this protein. The reduction in the number of residues that yield the conformational change in the holo→apo perturbation indicates the stabilization of the structure by the ligand. A similar observation was recently made for FBP (33). Thus, PRS may be used to locate functionally important residues in a given conformational change and it may also be used as a measure of conformational multiplicity of a given structure.

**ACKNOWLEDGEMENTS**. This work was partially supported by the Scientific and Technological Research Council of Turkey Project 106T522. SBO and ZNG acknowledge the Fulton High Performance Computing Initiative at ASU for computer time.

**Table 1. Results for proteins studied by PRS and ANM**

| | protein | apo/holo[1] | type of motion[2] | $N$[3] | rmsd(Å)[4] | $R_c^{opt}$(Å) (PRS/ANM)[5] | $M$[6] | PRS overlap (avg./best)[7] | PRS correlation (avg./best)[7] | ANM overlap (mode #)[8] | ANM correlation (mode #)[8] |
|---|---|---|---|---|---|---|---|---|---|---|---|
| subdomain motions | Thymidylate Synthase | 3tms/2tsc | shear | 264 | 0.80 | 8.6/8.6 | 1460 | 0.59 / 0.71 | 0.82 / 0.86 | 0.58 (1) | 0.82 (1) |
| | Tetracycline repressor (TR) | 1bjz/1bjy | shear | 179 | 0.83 | 12.1/11.0 | 2339 | 0.22 / 0.29 | 0.69 / 0.73 | 0.37 (16) | 0.77 (16) |
| | Small G protein Arf6 | 1e0s/2j5x | shear | 164 | 4.17 | 8.5/10.0 | 864 | 0.24 / 0.40 | 0.72 / 0.78 | 0.36 (10) | 0.64 (10) |
| | Annexin V | 1anx/1avr | hinge | 316 | 1.77 | 8.3/10.0 | 1635 | 0.42 / 0.56 | 0.64 / 0.70 | 0.26 (2) | 0.45 (2) |
| | FecA transporter | 1kmo/1kmp | hinge | 647 | 1.79 | 9.2/9.4 | 5042 | 0.39 / 0.54 | 0.72 / 0.75 | 0.37 (15) | 0.54 (15) |
| | OxyR transcription fact. | 1i6a/1i69 | not hinge or shear | 206 | 2.05 | 8.3/8.5 | 1012 | 0.17 / 0.32 | 0.83 / 0.88 | 0.24 (13) | 0.68 (2) |
| | Ubiquitin conj. enzyme (UCE) | 1j74/1j7d | not hinge or shear | 139 | 1.93 | 8.1/9.0 | 663 | 0.38 / 0.58 | 0.92 / 0.95 | 0.55 (2) | 0.88 (2) |
| domain motions | Cytochrome P450BM-3 | 1bu7/1jpz | shear | 453 | 1.13 | 7.9/8.1 | 2123 | 0.43 / 0.58 | 0.64 / 0.74 | 0.39 (8) | 0.74 (3) |
| | Molybdate-binding prot. | 1h9k/1h9m | shear | 141 | 0.87 | 10.2/10.2 | 1245 | 0.49 / 0.70 | 0.35 / 0.56 | 0.55 (2) | 0.44 (19) |
| | Adenylate Kinase (ADK) | 4ake/1ake | hinge | 214 | 7.13 | 8.0/8.0 | 984 | 0.79 / 0.92 | 0.95 / 0.96 | 0.80 (1) | 0.91 (1) |
| | Formate Dehydrogenase (FD) | 2nac/2nad | hinge | 374 | 1.18 | 8.0/8.0 | 1816 | 0.72 / 0.92 | 0.77 / 0.91 | 0.68 (3) | 0.52 (1) |
| | Ferric binding protein | 1d9v/1mrp | hinge | 309 | 2.48 | 7.7/8.0 | 1450 | 0.86 / 0.94 | 0.89 / 0.92 | 0.94 (1) | 0.90 (1) |
| | Maltose binding protein | 1omp/3mbp | hinge | 370 | 3.65 | 8.7/12.0 | 2341 | 0.85 / 0.95 | 0.76 / 0.91 | 0.86 (1) | 0.78 (2) |
| | Aldose reductase[4] | 2acq/1mar | not hinge or shear | 315 | 0.51 | 8.2/12.0 | 1593 | 0.41 / 0.58 | 0.81 / 0.86 | 0.54 (2) | 0.77 (1) |
| | Immunoglobulin | 1mcp/4fab | not hinge or shear | 214 | 5.90 | 7.5/8.0 | 942 | 0.68 / 0.73 | 0.55 / 0.62 | 0.65 (1) | 0.49 (1) |
| | HIV-1 Rev. Transcriptase | 2hmi/3hvt | partial refolding | 555 | 3.45 | 10.0/9.0 | 4592 | 0.64 / 0.77 | 0.47 / 0.59 | 0.51 (1) | 0.41 (4) |
| | Serpin | 1psi/7api | partial refolding | 372 | 8.60 | 9.5/9.0 | 2855 | 0.30 / 0.41 | 0.81 / 0.88 | 0.29 (3) | 0.75 (3) |
| subunit motions | Hemoglobin | 4hhb/2hco | allosteric | 141 | 0.73 | 9.3/11.0 | 971 | 0.40 / 0.54 | 0.77 / 0.80 | 0.46 (1) | 0.79 (3) |
| | VirB11 ATPase | 1g6o/1nlz | allosteric | 323 | 0.93 | 9.5/10.0 | 2299 | 0.63 / 0.79 | 0.69 / 0.71 | 0.67 (3) | 0.62 (4) |
| | Aspartate Receptor (AR) | 1lih/2lig | non-allosteric | 157 | 2.58 | 9.3/9.5 | 1025 | 0.36 / 0.53 | 0.74 / 0.87 | 0.37 (1) | 0.65 (12) |
| | MalK | 1q12/1q1b | non-allosteric | 367 | 1.16 | 10.5/8.0 | 3785 | 0.52 / 0.69 | 0.67 / 0.72 | 0.60 (2) | 0.63 (3) |
| | GroEL | 1aon/1oel | complex motions | 524 | 12.38 | 9.5/10.0 | 4393 | 0.80 / 0.87 | 0.86 / 0.92 | 0.80 (1) | 0.81 (1) |
| | Topoisomerase II (TII) | 1bgw/1bjt | complex motions | 664 | 20.57 | 12.0/13.0 | 8877 | 0.21 / 0.30 | 0.76 / 0.88 | 0.41 (7) | 0.75 (18) |

[1] PDB code

[2] as defined in the Yale Morph server (14)

[3] number of residues used in the analyses

[4] backbone RMS deviation except for aldose reductase where the **1mar** structure has only the $C_\alpha$ atoms available

[5] optimized cutoff distance as described under Methods for PRS/ANM analyses

[6] number of members in the network structure

[7] highest overlap (or Pearson correlation) between the scanned structures in the PRS technique and the experimental displacements, reported as the average over 10 runs as well as the single best result obtained from any residue.

[8] highest overlap (or Pearson correlation) between the eigenvectors of the apo structure and the experimental displacements; the best mode is reported in parentheses

**Table 2. Hot residues determined by PRS analysis**

| Transition | PDB Code | Hot Residues* |
|---|---|---|
| **GroEL** | | |
| T→R | 1AON(chain H)→2C7E(chain A) | 4-16,18,64-67,131, 137,174,190-192,374-376,410-418,474-476,486-494, 497,518-524 |
| R″→T | 1AON(chain A)→1GR5(chain A) | 100,103-104,107-108,146,149-151,156-157,159-160,184-185,197-207, 211-219,240-248,255-277,279-281,283-288,291,319,325-328,343-369,372 |
| **ADK** | | |
| Open→Closed | 4AKE→1AKE | 5-10,13, 17, 30-57, 59, 62-73, 84-85, 94-101, 106-110, 120-133, 140-156, 158-159, 167-178, 180-187, 193, 196-197 |
| Closed→Open | 1AKE→4AKE | 53-56, 125, 152, 154, 158 |

*Residues that yield Pearson correlations higher than 0.7 in PRS analysis (averaged over 10 runs).

**FIGURE LEGENDS**

**Fig 1.** $C_\alpha$ displacement results for (A) ADK, (B) AR, (C) UCE, and (D) FD. Gray line represents the experimental $C_\alpha$ displacements between open and closed structures, black curve is the prediction of PRS, dashed curve is obtained from the best normal mode vector in ANM. Area under each curve is normalized to unity. Respective correlation between the experimental conformational change and theoretical binding induced fluctuation profiles of PRS and ANM are 0.96 and 0.91 for ADK, 0.87 and 0.65 for AR, 0.95 and 0.88 for UCE and 0.91 and 0.52 for FD. Insets show ribbon diagrams of superimposed proteins; open form is in black and closed form is in gray. PRS performs better than ANM slow modes in predicting the binding induced conformational changes especially for proteins in C and D.

**Fig 2.** (A) Average redundancy index vs the variance plot for the proteins listed in Table 1, colored according to the mode that best represents the conformational change between the apo and holo forms. Experimental binding induced conformational change in proteins with low redundancy index (and low variance) are well-represented by one of the four slowest modes. (B) Average path length vs average redundancy per residue and (C) Probability distribution of redundancy index for the residues of ADK (black) and hemoglobin (red).

**Fig 3**. Ribbon diagrams of GroEL colored according to correlation coefficients between the theoretical and experimental conformational change of T→R transition (A) and R″→T transition (B). The residues whose perturbation leads to response vector profiles highly correlated with experimental conformational changes are red, whereas residues with the least correlated profile are shown in blue within a color spectrum of red-orange-yellow-green-cyan and blue. Residues with the highest correlation ($C_i > 0.69$) are shown in ball representation.

**Fig 4.** Ribbon diagrams of ADK: (A) apo form, (B) holo form crystallized with inhibitor AP$_5$A. Each residue in the ribbon diagram is colored according the correlation between the experimental binding induced conformational change and the theoretical response profile upon perturbing that residue, using a spectrum of red (for highest correlation) to orange, yellow, green cyan and blue (lowest correlation). The perturbed residues with the highest correlations are shown as balls. Also displayed are the color-coded normalized $\mathbf{H}^{-1}$ matrices for open (C) and closed states (D). Normalization is carried out so that the displacement in each direction of every residue is equal.

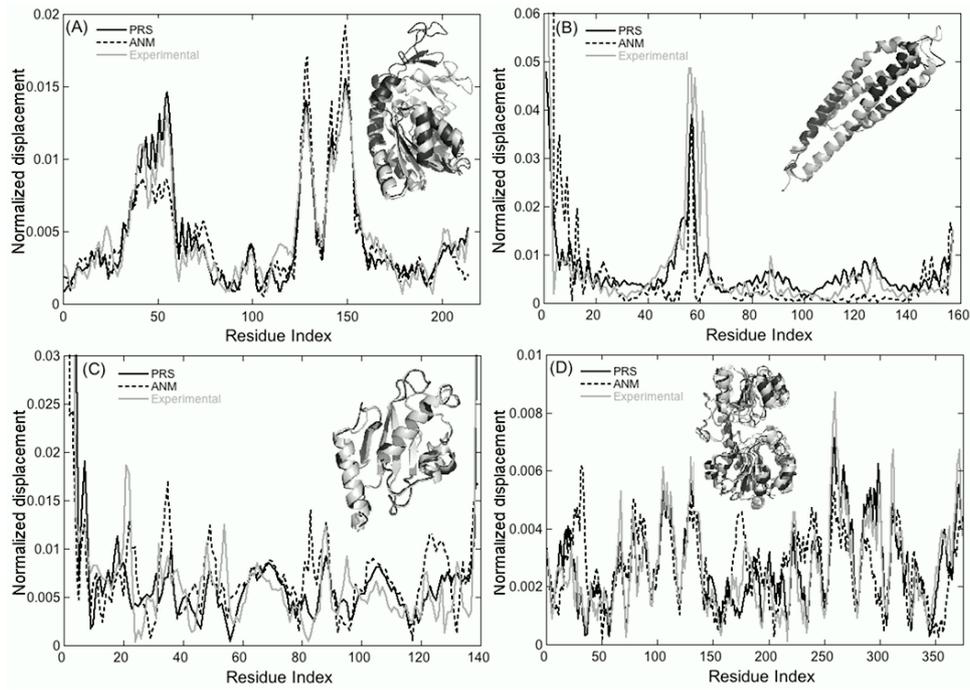

**Figure 1**

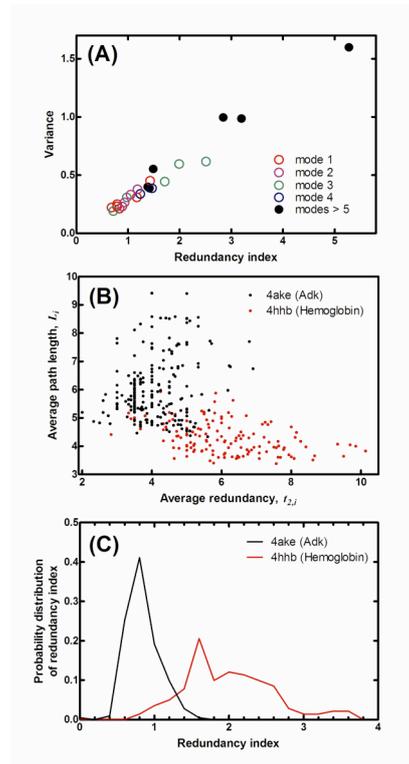

**Figure 2**

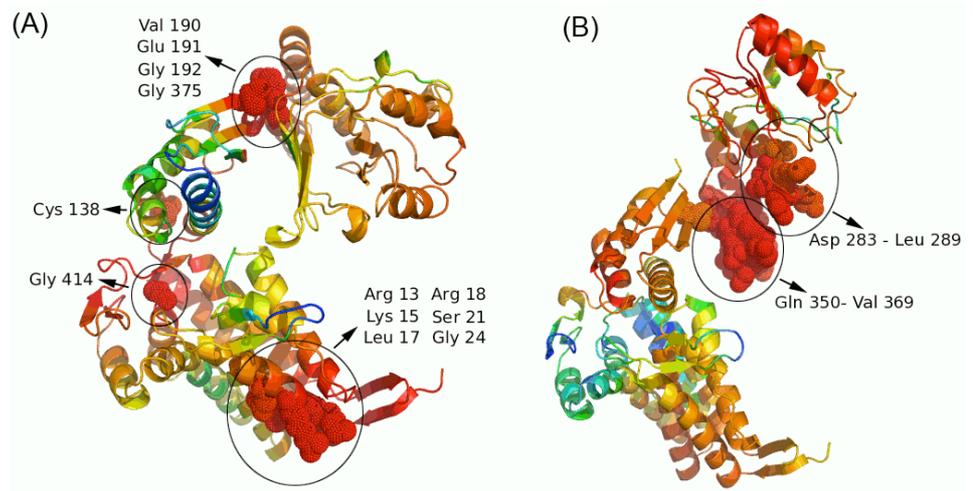

**Figure 3**

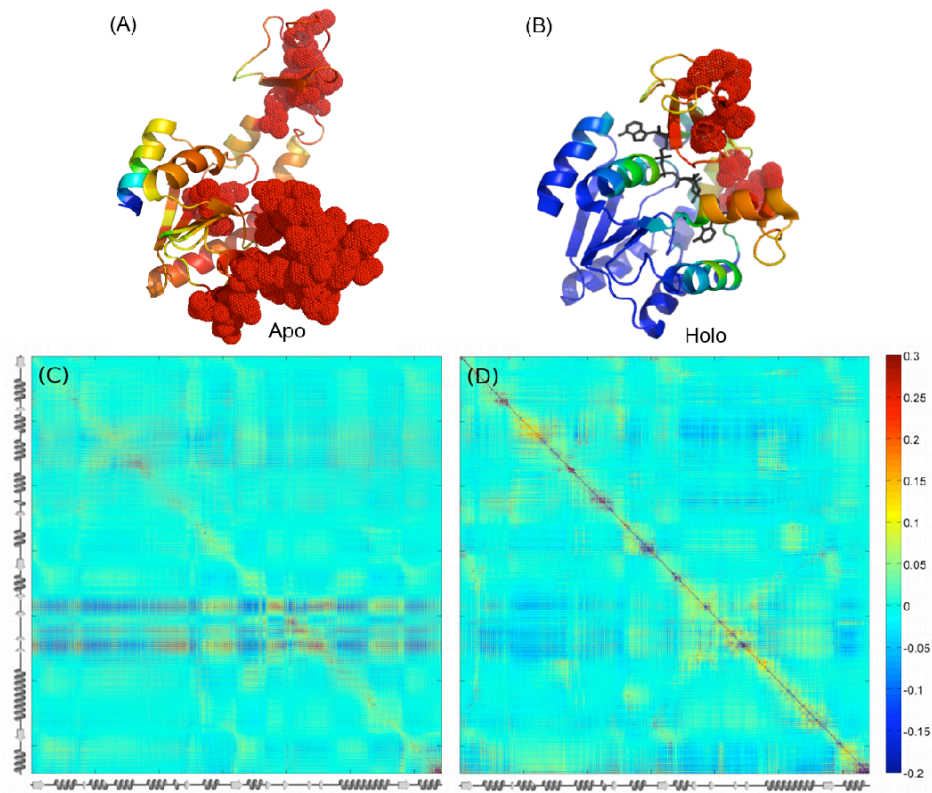

**Figure 4**